\documentclass[3p,times,twocolumn]{elsarticle}

\usepackage{ecrc}

\volume{00}

\firstpage{1}

\journalname{Nuclear Physics B Proceedings Supplement}

\runauth{}

\jid{nuphbp}

\jnltitlelogo{Nuclear Physics B Proceedings Supplement}

\usepackage{amssymb}

\usepackage[figuresright]{rotating}

\begin{document}

\begin{frontmatter}



\dochead{}

\title{Learning about the CP phase in the next 10 years}


\author{Pedro A. N. Machado}

\address{Departamento de F\'isica Te\'orica and Instituto de
  F\'{\i}sica Te\'orica, IFT-UAM/CSIC,\\ Universidad Aut\'onoma de
  Madrid, Cantoblanco, 28049, Madrid, Spain}

\begin{abstract}
We assess the sensitivity to the lepton CP phase by accelerator and
reactor experiments in the near future, characterizing it globally by
means of the CP exclusion fraction measure. Such measure quantifies
what fraction of the $\delta_{\rm CP}$ space can be excluded at given
input values of $\theta_{23}$ and $\delta_{\rm CP}$. For some region
of the parameter space, we find that T2K and NO$\nu$A combined can
exclude about $30\%-40\%$ of the $\delta_{\rm CP}$ space at $3\sigma$
with a 5 years running in each neutrino and antineutrino modes.  A
determination of the mass hierarchy would be possible for a modest
portion of the parameter space at $3\sigma$.
\end{abstract}


\end{frontmatter}


\section{Introduction}
\label{introduction}
After the discovery of $\theta_{13}$~\cite{Abe:2011sj, Abe:2011fz,
  An:2012eh, Ahn:2012nd}, the attention has been turned upon the last
two unknowns in the three neutrino oscillation framework, the lepton
CP violating phase $\delta_{\rm CP}$ and the neutrino mass
hierarchy. Gathering any information on these two unknowns is highly
valuable. The determination of the mass hierarchy combined with the
observability or not of neutrinoless double beta decay might reveal
the Dirac or Majorana nature of neutrinos. In such scenarios, one
might envisage that future sensitivities on the sum of neutrino masses
from cosmological data could provide a powerful tool to test the
standard cosmology framework.  In this work, we focus on how much
information on $\delta_{\rm CP}$ can be obtained by current neutrino
oscillation experiments, namely T2K and NO$\nu$A, as how this
correlates with the mass hierarchy measurement.

Certainly, future facilities such as
Hyper-Kamiokande~\cite{Abe:2011ts} and LBNE~\cite{Akiri:2011dv} are
supposed to have remarkable sensitivities to both hierarchy and lepton
CP violation~\cite{Blennow:2014sja}. Nevertheless, given the time
scale of such experiments, any knowledge of these quantities is
valuable.  The main goal of this manuscript is to show under which
circumstances T2K and NO$\nu$A, are able to exclude a significant
fraction of the $\delta_{\rm CP}$ values.


\section{The appropriate CP measure for T2K and NO$\nu$A}
\label{CP-ex-fraction}

In this paper, to quantify the experimental sensitivity of T2K and
NO$\nu$A to $\delta_{\rm CP}$ in a global way, we use the \emph{CP
  exclusion fraction}, $f_{\rm CPX}$, where $f_{\rm CPX}$ is defined as
the fraction of $\delta_{\rm CP}$ values which can be disfavored at a
given confidence level for a given set of input
parameters~\cite{Winter:2003ye,Huber:2004gg}.

One may be tempted to say that choosing a measure to assess the CP
sensitivity is to a large extent arbitrary. Ultimately, such assertion
is indeed true. Nonetheless, we argue that for experiments like T2K
and NO$\nu$A, $f_{\rm CPX}$ is more useful than other routinely used
measures. The CP violation fraction, for instance, gives us the
fraction of $\delta_{\rm CP}$ values for which CP violation can be
established (see, e.g., Refs.~\cite{Abe:2011ts,Huber:2009cw}). Regardless
the straightforward binary ``yes or no'' message to CP violation, this
measure suffers from the ``bias'' of choosing special reference points
($\delta_{\rm CP} = 0$ or $\pi$) to discuss the sensitivity to CP
phase, losing valuable information: even if the experiments
aforementioned cannot establish CP violation, they may still disfavor
some region of the parameter space.  

The expected error in $\delta_{\rm CP}$, on the other hand, suffers
from a conceptual problem when dealing with T2K and NO$\nu$A.  The
interpretation of the allowed region as an error bar fails if such
region consists of disconnected pieces or if there are multiple fake
solutions and non-Gaussianities in the $\chi^2$~\cite{Huber:2004gg}.

Probably the most straightforward way of showing the sensitivity to
$\delta_{\rm CP}$ is to plot the allowed region in, say, the
$\delta_{\rm CP}\times\sin^2\theta_{23}$ plane for many different
input parameters. The issue for the experiments we are interested in
is that even the $1\sigma$ allowed regions are very large, sometimes
disconnected, making it hard to build up a global perspective of the
situation.

Hence, we advocate that the most appropriate way of evaluating
globally the CP sensitivity of these experiments is to make use of the
$f_{\rm CPX}$. We will present our results in the plane $\delta_{\rm
  CP}\times\sin^2\theta_{23}$, as $\theta_{23}$ now is the least
constrained angle, affecting considerably the experimental outcome.
In this work, we use the standard parameterization of the neutrino
mixing matrix \cite{Beringer:1900zz}.

\section{Results}
\label{results}

\begin{figure}[!h]
\begin{center}
T2K$+$NO$\nu$A: $5+5$ years each -- inverted hierarchy
\includegraphics[width=0.48\textwidth]{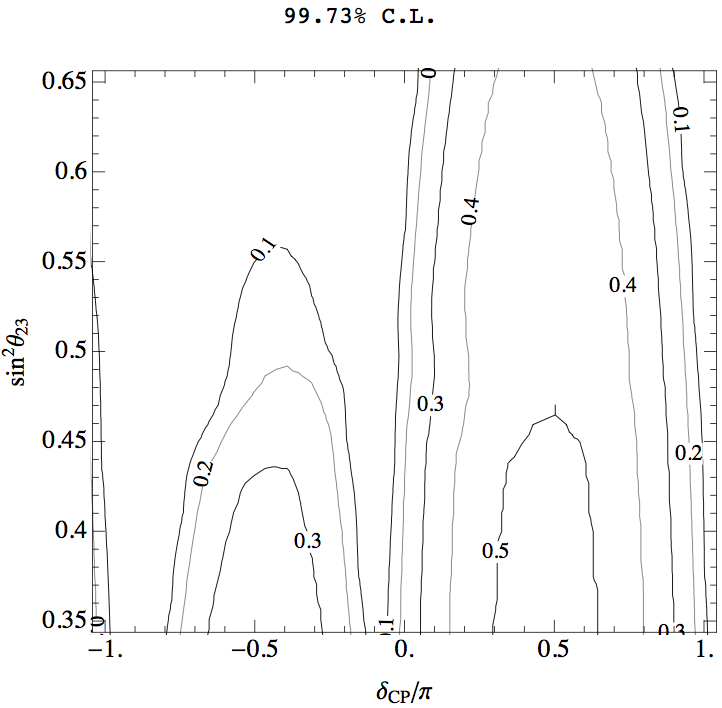}
Fitting only normal hierarchy
\includegraphics[width=0.48\textwidth]{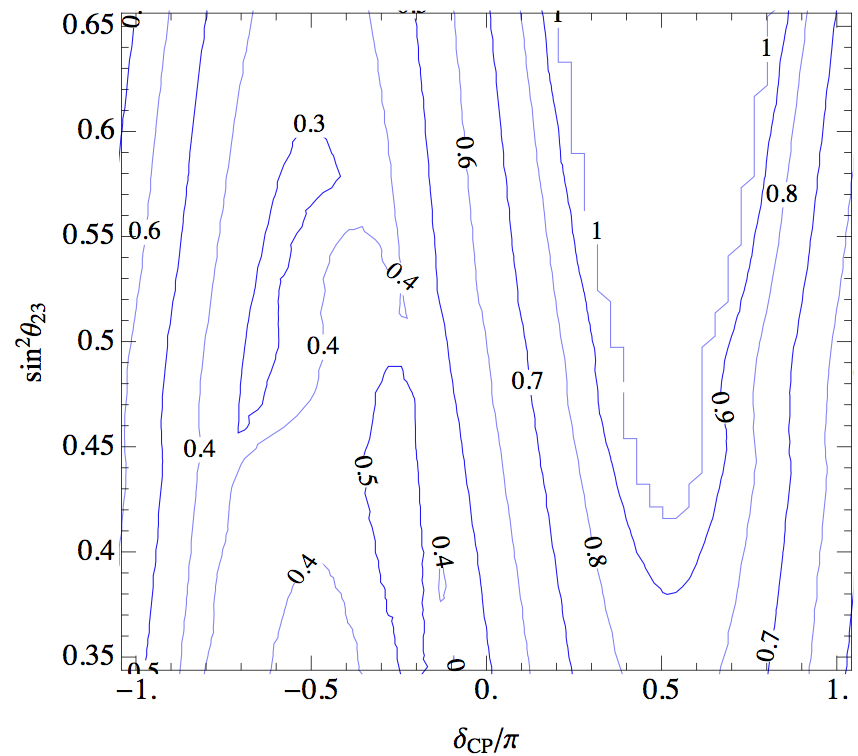}
\end{center}
\vspace{-7mm}
\caption{$f_{\rm CPX}$ isolines on the $\delta_{\rm CP} - \sin^2
  \theta_{23}$ plane at $3\sigma$, for T2K and NO$\nu$A running in
  $\nu$+$\bar \nu$ mode for $5+5^\prime$ years, assuming inverted
  hierarchy as input. On the top, we marginalized over the hierarchy,
  while on the bottom we fit only the normal hierarchy (that is, the
  wrong hierarchy in this case).}
\label{3sig}
\end{figure}

In Fig.~\ref{3sig}, we present the $\delta_{\rm CP}$ sensitivity for
the combination of T2K and NO$\nu$A, assuming a running of 5 years in
each neutrino and antineutrino mode for both experiments. We take into
account that $\theta_{13}$ is expected to be measured by Daya Bay with
an error of $\sim 0.005$. The details of the simulations can be found in
Ref.~\cite{Machado:2013kya}, as well as a qualitative discussion based
on the bi-probability plots and on spectral information.

Very roughly, the $f_{\rm CPX}$ contours for normal as input can be
obtained by reparametrizing $\delta_{\rm CP} \to \pi-\delta_{\rm CP}$
in Fig.~\ref{3sig}. We show on the top panel the result marginalizing
over the hierarchy. We can see that generically, as
$\sin^2\theta_{23}$ grows, the sensitivities are worsened. When
$\theta_{23}$ goes to the second octant, it boosts the leading (CP
conserving) $\nu_\mu\to\nu_e$ oscillation term, inducing an increased
number of $\nu_e$ appearance events. Although this improves the
sensitivity to $\theta_{13}$, such events are a \emph{background} to
the CP phase search, deteriorating $f_{\rm CPX}$. 

Another salient feature is the disparate performance for positive and
negative values of $\delta_{\rm CP}$. What happens here can be
understood by the degeneracy between $\delta_{\rm CP}$ and the
hierarchy~\cite{Machado:2013kya}: in this region, the wrong hierarchy
is likely to provide a very good fit of the data for values of
$\delta_{\rm CP}$ which would be disfavored in the correct hierarchy
fit. This bring us to the bottom panel in Fig.~\ref{3sig}, where we
fit the expected data with only the normal hierarchy, that is,
\emph{the wrong hierarchy}.

On the left side of the plane, most of the wrong hierarchy
$\delta_{\rm CP}$ values can fit the data, displaying the degeneracy
between $\delta_{\rm CP}$ and the hierarchy.  This reduces the $f_{\rm
  CPX}$ to the point that it becomes zero (at $3\sigma$) for a large
portion of the input parameters. On the other side of the parameter
space, the degeneracy is lifted mainly due to the matter effects in
NO$\nu$A, while the final size of the allowed region of $\delta_{\rm
  CP}$ is decisively impacted by the performance of T2K -- an evidence
for the synergy between these experiments. For maximal values of
$\delta_{\rm CP}$, there is the possibility of even rejecting the
wrong hierarchy at $3\sigma$. Although this sounds very promising, we
call the attention that such a scenario is restricted to a fairly
small fraction of the parameter space.

A few comments are in order regarding the $\theta_{23}$ octant
determination.
The octant degeneracy has its roots in the dependency of the
disappearance mode $\nu_\mu\to\nu_\mu$ and its CPT conjugate in
$\sin^22\theta_{23}$. By themselves, these modes cannot determine
the octant of the atmospheric mixing angle. The appearance mode
$\nu_\mu\to\nu_e$ and its CP conjugate, on the other hand, have an
intricate dependency on $\delta_{\rm CP}$, $\sin^2\theta_{23}$, and
on the hierarchy, and they are plagued by
degeneracies~\cite{Hiraide:2006vh}.
The inclusion of antineutrino running is essential to break these
degeneracies, being an indispensable ingredient for both the CP phase
the octant~\cite{Machado:2013kya} sensitivities. The issue is the loss
of statistics when running in the antineutrino mode due to smaller
fluxes and cross sections.  We have checked that the octant
sensitivity does not change much if the antineutrino share is between
30\% to 70\% of the total running time. This, together with the
importance of the determination of $\delta_{\rm CP}$ and the
hierarchy, indicates that the octant determination should play a minor
role in the experiments run planning.

\section{Conclusions}
\label{conclusions}
In the next 10 years, we do not expect a precise measurement of the
lepton CP phase, as the present neutrino oscillation experiments have
very limited statistics for such goal. Notwithstanding, given the
large value of $\theta_{13}$, these experiments may be able to provide
valuable information on the phase. Depending on the true values of
$\delta_{\rm CP}$ and $\theta_{23}$, they might be able to strongly
disfavor a reasonable region of the parameter space.

We have
quantified the expected sensitivity of T2K and NO$\nu$A by using the
CP exclusion fraction -- the fraction of the CP values that can be
excluded for a given input at a given confidence level (we adopted
$3\sigma$ in this paper).
We have shown that, for some region of the true parameter space, about
$30\%-50\%$ of the $\delta_{\rm CP}$ space can be strongly disfavored
with a 5 years running in each neutrino and antineutrino modes. This
is because of the synergy between T2K and NO$\nu$A, as the latter is
more sensitive to the hierarchy, due to a longer baseline, while the
first is more sensitive to the phase, once the wrong hierarchy is
disfavored. In fact, there is even a modest possibility of determining
the neutrino mass hierarchy at $3\sigma$ in the next 10 years.

The use of the antineutrino appearance channel is crucial for
obtaining information on $\delta_{\rm CP}$, the optimal situation
being the equal share between neutrino and antineutrino runnings.

\section{Acknowledgements}
The work of PANM is supported by an ESR contract of the European Union
network FP7 ITN INVISIBLES (Marie Curie Actions,
PITN-GA-2011-289442). 




\nocite{*}
\bibliographystyle{elsarticle-num}
\bibliography{nuphbp-template}

\begin{thebibliography}{10}
\expandafter\ifx\csname url\endcsname\relax
  \def\url#1{\texttt{#1}}\fi
\expandafter\ifx\csname urlprefix\endcsname\relax\def\urlprefix{URL }\fi
\expandafter\ifx\csname href\endcsname\relax
  \def\href#1#2{#2} \def\path#1{#1}\fi

\bibitem{Abe:2011sj}
K.~Abe, et~al., {Indication of Electron Neutrino Appearance from an
  Accelerator-produced Off-axis Muon Neutrino Beam}, Phys.Rev.Lett. 107 (2011)
  041801.
\newblock \href {http://arxiv.org/abs/1106.2822} {\path{arXiv:1106.2822}},
  \href {http://dx.doi.org/10.1103/PhysRevLett.107.041801}
  {\path{doi:10.1103/PhysRevLett.107.041801}}.

\bibitem{Abe:2011fz}
Y.~Abe, et~al., {Indication for the disappearance of reactor electron
  antineutrinos in the Double Chooz experiment}, Phys.Rev.Lett. 108 (2012)
  131801.
\newblock \href {http://arxiv.org/abs/1112.6353} {\path{arXiv:1112.6353}},
  \href {http://dx.doi.org/10.1103/PhysRevLett.108.131801}
  {\path{doi:10.1103/PhysRevLett.108.131801}}.

\bibitem{An:2012eh}
F.~An, et~al., {Observation of electron-antineutrino disappearance at Daya
  Bay}, Phys.Rev.Lett. 108 (2012) 171803.
\newblock \href {http://arxiv.org/abs/1203.1669} {\path{arXiv:1203.1669}},
  \href {http://dx.doi.org/10.1103/PhysRevLett.108.171803}
  {\path{doi:10.1103/PhysRevLett.108.171803}}.

\bibitem{Ahn:2012nd}
J.~Ahn, et~al., {Observation of Reactor Electron Antineutrino Disappearance in
  the RENO Experiment}, Phys.Rev.Lett. 108 (2012) 191802.
\newblock \href {http://arxiv.org/abs/1204.0626} {\path{arXiv:1204.0626}},
  \href {http://dx.doi.org/10.1103/PhysRevLett.108.191802}
  {\path{doi:10.1103/PhysRevLett.108.191802}}.

\bibitem{Abe:2011ts}
K.~Abe, T.~Abe, H.~Aihara, Y.~Fukuda, Y.~Hayato, et~al., {Letter of Intent: The
  Hyper-Kamiokande Experiment --- Detector Design and Physics Potential
  ---}\href {http://arxiv.org/abs/1109.3262} {\path{arXiv:1109.3262}}.

\bibitem{Akiri:2011dv}
T.~Akiri, et~al., {The 2010 Interim Report of the Long-Baseline Neutrino
  Experiment Collaboration Physics Working Groups}\href
  {http://arxiv.org/abs/1110.6249} {\path{arXiv:1110.6249}}.

\bibitem{Blennow:2014sja}
M.~Blennow, P.~Coloma, E.~Fernandez-Martinez, {Reassessing the sensitivity to
  leptonic CP violation}, JHEP 1503 (2015) 005.
\newblock \href {http://arxiv.org/abs/1407.3274} {\path{arXiv:1407.3274}},
  \href {http://dx.doi.org/10.1007/JHEP03(2015)005}
  {\path{doi:10.1007/JHEP03(2015)005}}.

\bibitem{Winter:2003ye}
W.~Winter, {Understanding CP phase dependent measurements at neutrino
  superbeams in terms of bi-rate graphs}, Phys.Rev. D70 (2004) 033006.
\newblock \href {http://arxiv.org/abs/hep-ph/0310307}
  {\path{arXiv:hep-ph/0310307}}, \href
  {http://dx.doi.org/10.1103/PhysRevD.70.033006}
  {\path{doi:10.1103/PhysRevD.70.033006}}.

\bibitem{Huber:2004gg}
P.~Huber, M.~Lindner, W.~Winter, {From parameter space constraints to the
  precision determination of the leptonic Dirac CP phase}, JHEP 0505 (2005)
  020.
\newblock \href {http://arxiv.org/abs/hep-ph/0412199}
  {\path{arXiv:hep-ph/0412199}}, \href
  {http://dx.doi.org/10.1088/1126-6708/2005/05/020}
  {\path{doi:10.1088/1126-6708/2005/05/020}}.

\bibitem{Huber:2009cw}
P.~Huber, M.~Lindner, T.~Schwetz, W.~Winter, {First hint for CP violation in
  neutrino oscillations from upcoming superbeam and reactor experiments}, JHEP
  0911 (2009) 044.
\newblock \href {http://arxiv.org/abs/0907.1896} {\path{arXiv:0907.1896}},
  \href {http://dx.doi.org/10.1088/1126-6708/2009/11/044}
  {\path{doi:10.1088/1126-6708/2009/11/044}}.

\bibitem{Beringer:1900zz}
J.~Beringer, et~al., {Review of Particle Physics (RPP)}, Phys.Rev. D86 (2012)
  010001.
\newblock \href {http://dx.doi.org/10.1103/PhysRevD.86.010001}
  {\path{doi:10.1103/PhysRevD.86.010001}}.

\bibitem{Machado:2013kya}
P.~Machado, H.~Minakata, H.~Nunokawa, R.~Zukanovich~Funchal, {What can we learn
  about the lepton CP phase in the next 10 years?}, JHEP 1405 (2014) 109.
\newblock \href {http://arxiv.org/abs/1307.3248} {\path{arXiv:1307.3248}},
  \href {http://dx.doi.org/10.1007/JHEP05(2014)109}
  {\path{doi:10.1007/JHEP05(2014)109}}.

\bibitem{Hiraide:2006vh}
K.~Hiraide, H.~Minakata, T.~Nakaya, H.~Nunokawa, H.~Sugiyama, et~al.,
  {Resolving $\theta_{23}$ degeneracy by accelerator and reactor neutrino
  oscillation experiments}, Phys.Rev. D73 (2006) 093008.
\newblock \href {http://arxiv.org/abs/hep-ph/0601258}
  {\path{arXiv:hep-ph/0601258}}, \href
  {http://dx.doi.org/10.1103/PhysRevD.73.093008}
  {\path{doi:10.1103/PhysRevD.73.093008}}.

\end{thebibliography}







\end{document}